\documentclass[showpacs, twocolumn]{revtex4-1}
\usepackage{graphicx}
\usepackage{amsmath}
\usepackage{appendix}
\begin{document}

\title{Superfluidity and Bose-Einstein condensation in a dipolar Bose gas with weak disorder}

\author{Abdel\^{a}ali Boudjem\^{a}a}

\affiliation{Department of Physics, Faculty of Sciences, Hassiba Benbouali University of Chlef P.O. Box 151, 02000, Ouled Fares, Chlef, Algeria.}

%\date{\today}

\begin{abstract}
We investigate the properties of  a  three-dimensional  homogeneous dipolar Bose gas in a weak random potential with a Gaussian correlation function at finite temperature. 
Using the Bogoliubov theory (beyond the mean field), we calculate the superfluid and the condensate fractions 
in terms of the interaction strength on the one hand and in terms of the width and the strength of the disorder on the other.
The influence of the disordered potential on the second order correlation function, the ground state energy and the chemical potential is also analyzed. 
We find that for fixed strength and correlation length of the disorder potential, the dipole-dipole interaction leads to modify both the condensate and the superfluid fractions. 
We show that for a strong disorder strength the condensed fraction becomes larger than the superfluid fraction. 
We discuss the effect of the trapping potential on a disordered dipolar Bose in the regime of large number of particles.
\end{abstract}


\maketitle

\section{Introduction}
Recently, the study of Bose-Einstein condensate (BEC) in disordered potentials, known as "dirty boson problem" \cite{Fish}, has paid a special attention.
Experimentally, it was first studied in the context of the motion of superfluid helium in porous Vycor glass \cite {Chan}. 
%Recently, the physics of dilute Bose gases  in random potentials has attracted a considerable attention both theoretically and experimentally.
The problem of boson localization, the superfluid-insulator transition, and the nature of elementary
excitations in lattice models with random site potentials characterized by Hubbard or equivalent models have been the object of several theoretical investigations  \cite {Ma, Sing} 
and Monte Carlo numerical simulations \cite {Li, Krauth, Zhang}. Such random potentials suppress or may even
completely destroy the long-range order related to BEC \cite{Anders} leading to the transition to the so-called Bose glass phase \cite{Fish}. 
This state is characterized by a finite compressibility, the absence of a gap in the single particle spectrum, and a nonvanishing density of states at zero energy.
Bose systems with disorder have also been addressed in the uniform system without lattices i.e. continuum regime
using the Bogoliubov model \cite {Huang, Mich}, hydrodynamic approximation\cite {Gior}, Beliaev-Popov diagrammatic technique \cite {Lopa, Axel}, mean-field theory \cite {Yuk1},
and Monte Carlo simulations \cite {Gior1}.
The main finding of the above studies is that both BEC and superfluidity are depressed due the competition between the two-body repulsive interaction and the random potential.
The random potential can be created using different techniques for example, laser speckles which are produced by focusing the laser beam
on a glass plate \cite {Clem, Abdulaev}, wire traps represent magnetic traps on atomic chips \cite {Krug}. Also,  deep  \cite {Cad} and incommensurable lattices provide a
useful random medium\cite {Asp1}.

On the other hand, there has been immense progress in creating BECs with strong magnetic dipoles \cite {Pfau, Pupillo2012, Lu, Aik, Lu1}, 
and heteronuclear molecules with strong electric dipoles \cite { Aik1, kk, Aik2}  (e.g., RbCs) as well as Rydberg atoms in electric fields \cite {Tak}.
What renders such systems particularly fascinating is that the atoms interact via a dipole-dipole interaction (DDI) that is both long-ranged and anisotropic 
which leads to the observation of novel phenomena in ultracold atomic gases \cite {Pfau, Pupillo2012, Baranov}. 
Recently, uniform dipolar Bose gas with a Gaussian disorder correlation function \cite {Axel1} 
and a Lorentzian \cite {Axel2}, have been investigated using mean field theory at zero temperature.
Most recently, a delta-correlated disorder is examined by the same group \cite {Axel3} employing the Bogoliubov theory at finite temperatures.
The main result emerging from these explorations is that the anisotropy of the two-particle DDI is passed on to both BEC and the superfluid fraction.

%In this paper, we extend the work of Ref \cite {Axel1} at finite temperature, such a regime remains poorly explored in theory and experiments.
In this paper, we study the properties of a three-dimensional (3D) homogeneous dipolar Bose gas in a weak random potential with Gaussian correlation 
by employing a perturbation theory at finite temperature.
%such a regime remains poorly explored in theory and experiments. 
The Gaussian-correlated disorder is defined by both the strength and the correlation length of the disorder (see Eq.(\ref{Disp}))  which permit
us to well control the interplay between the disorder correlation length and the interaction strength. 
Therefore, this makes our model completely different from that of Ref.\cite {Axel3} and thus, our results are also different in particular at zero temperature.
Among the advantages of the Gaussian-correlated disorder potential is that its presence in 1D Bose gas with 
short-range potential can undergo a finite-temperature phase transition between two distinct states: fluid and insulator \cite{Alts}.
In 3D case, BEC with a weak, Gaussian-correlated disordered potential can constitute a good model to describe the Bose glass phase \cite {Mich}.
%In 2D Bose gas with DDI, one can expect that such a random potential may yield the transition into a superglass state \cite{Boudj}.\\
Moreover, we investigate, for the first time, the behavior of the anomalous density of a dirty BEC with DDI. This quantity which grows with interactions and vanishes
in noninteracting systems\cite {Griffin, boudj2010, boudj2011}, is important to fully understand the interplay of disorder and interactions.
The anomalous density is often neglected in the literature without any physical or mathematical justifications. Indeed, it is usually of order or even larger than the noncondensed density 
(see Eq.(\ref{heis})) regardless of the system in which it is studied namely BEC with pure contact interaction \cite {Griffin,Burnet, Yuk, boudj2010, boudj2011, boudj2012}, 
BEC-impurity mixtures\cite {boudj2014, boudjA2014} or dipolar BEC \cite {boudj2015}. 
Moreover, the absence of the anomalous averages renders the system unstable.
We show, in addition, how the anisotropy of the DDI enhances quantum, thermal and disorder fluctuations as well as the superfluid fraction.
We find that the thermal fluctuations' contribution to the condensate depletion and the superfluid fraction coincide with those  obtained in \cite {Axel3} for a delta-correlated disorder. 
This is in fact natural since we assume that disorder, quantum and thermal fluctuations are too small that the Hamiltonian of the system can be expanded in 
both leading order of disorder plus thermal and quantum fluctuations. Therefore, disorder and thermal fluctuations are additive and, thus, independent from each other.
%It is to be noted here that these terms depend only on the DDI and remain the same for any disorder correlation function namely : 
%delta function, Gauss function, laser speckles, or a Lorentzian. 

The rest of the paper is organized as follows. In section \ref {model}, we describe our model of the dipolar dilute Bose gas in a general random potential.
In section \ref{zeroT}, we derive analytical expressions for the condensate fluctuations, the second order correlation function and some 
thermodynamic quantities for external random potential with Gaussian correlation at finite temperature
and compare our finding with the mean field results of Ref \cite{Axel1}.
We show that the competition between both contact interaction-disorder and DDI-disorder leads to enhance the condensate depletion, 
the anomalous density, disorder fluctuation and the ground state energy. 
We establish the validity criteria of the Bogoliubov theory of disordered dipolar systems.
In section \ref{superfluid}, the superfluid fraction is obtained and its characteristics are discussed.
In section \ref{Trap},  we extend our results to the case of a harmonically trapped gas by making use of the local density approximation.
Finally, our conclusions remain in section\ref{conc}.

\section{The model} 
\label {model}
We consider the effects of an external random field on a dilute 3D dipolar Bose gas with dipoles oriented perpendicularly to the plane.
The Hamiltonian of the system is written as:
\begin{align}\label{ham}
&\hat H = \int d^3r \, \hat \psi^\dagger ({\bf r}) \left(\frac{-\hbar^2 }{2m}\Delta+U({\bf r})\right)\hat\psi(\mathbf{r}) \nonumber \\
&+\frac{1}{2}\int d^3r\int d^3r^\prime\, \hat\psi^\dagger(\mathbf{r}) \hat\psi^\dagger (\mathbf{r^\prime}) V(\mathbf{r}-\mathbf{r^\prime})\hat\psi(\mathbf{r^\prime}) \hat\psi(\mathbf{r}) ,
\end{align}
where $\psi^\dagger$ and $\psi$ denote, respectively the usual creation and annihilation field operators, the interaction potential
$V(\mathbf{r}-\mathbf{r^\prime})=g\delta(\mathbf{r}-\mathbf{r^\prime})+V_{dd}(\mathbf{r}-\mathbf{r^\prime})$, $g=4\pi \hbar^2 a/m$
corresponds to the short-range part of the interaction and is parametrized by the scattering length as $a$. 
It is important to mention here that the contact potential exhibits ultraviolet divergences in both 2D and 3D geometries.
Common ways to cure these inconsistencies is the use of either renormalization or a pseudopotential. \\
On the other hand, the dipole-dipole component reads  
\begin{equation}\label{dd}
V_d({\bf r}) =\frac{C_{dd}}{4\pi}\frac{1-3\cos^2\theta}{r^3},
\end{equation}
where the coupling constant $C_{dd} $ is $M_0 M^2$ for particles having a permanent magnetic dipole moment $M$ ($M_0$ is the magnetic permeability
in vacuum) and $d^2/\epsilon_0$ for particles having a permanent electric dipole $d$ ($\epsilon_0 $ is the permittivity of vacuum),
$m$ is the particle mass, and $\theta$ is the angle between the relative position of the particles $\vec r$ and the direction of the dipole.
The characteristic dipole-dipole distance can be defined as $r_*=m C_{dd}/4\pi \hbar^2$. %For most polar molecules $r_*$ ranges from 10 to $10^4$ \AA. 
The disorder potential is described by vanishing ensemble averages $\langle U(\mathbf r)\rangle=0$
and a finite correlation of the form $\langle U(\mathbf r) U(\mathbf r')\rangle=R (\mathbf r,\mathbf r')$.

Passing to the Fourier transform and working in the momentum space, the Hamiltonian (\ref {ham}) takes the form:
\begin{align}\label{ham1}
&\hat H\!\!=\!\!\sum_{\bf k}\! \frac{\hbar^2k^2}{2m}\hat a^\dagger_{\bf k}\hat a_{\bf k}\! +\!\frac{1}{V}\!\!\sum_{\bf k,\bf p} \! U_{\bf k\!-\!\bf p} \hat a^\dagger_{\bf k} \hat a_{\bf p}  \\ \nonumber
&+\!\frac{1}{2V}\!\!\sum_{\bf k,\bf q,\bf p}\!\!
f ({\bf p})\hat a^\dagger_{\bf k\!+\!\bf q} \hat a^\dagger_{\bf k\!-\!\bf q}\hat a_{\bf k\!+\!\bf p}\hat a_{\bf k\!-\!\bf p} ,
\end{align}
where $V$ is a quantization volume, and the interaction potential in momentum space is given by \cite{boudj2015}
\begin{equation}\label{scam}
 f(\mathbf k)=g[1+\epsilon_{dd} (3\cos^2\theta_k-1)],
\end{equation}
here $\epsilon_{dd}=C_{dd} /3g$ is the dimensionless relative strength which describes the interplay between the DDI and short-range interactions.\\
Assuming the weakly interacting regime where  $r_*\ll \xi$ with $\xi=\hbar/\sqrt{mgn}$ being the healing length and $n$ is the total density,
we may use the Bogoliubov approach. Applying the inhomogeneous Bogoliubov transformations \cite{Huang}: 
\begin{equation}\label {trans}
 \hat a_{\bf k}= u_k \hat b_{\bf k}-v_k \hat b^\dagger_{-\bf k}-\beta_{\bf k},  \qquad \hat a^\dagger_{\bf k}= u_k \hat b^\dagger_{\bf k}-v_k \hat b_{-\bf k}-\beta_{\bf k}^*,
\end{equation}
where $\hat b^\dagger_{\bf k}$ and $\hat b_{\bf k}$ are operators of elementary excitations.
The Bogoliubov functions $ u_k,v_k$ are expressed in a standard way:
$ u_k,v_k=(\sqrt{\varepsilon_k/E_k}\pm\sqrt{E_k/\varepsilon_k})/2$ with $E_k=\hbar^2k^2/2m$ is the energy of free particle, and 
\begin{equation}\label {beta}
\beta_{\bf k}=\sqrt{\frac{n}{V}} \frac{E_k}{\varepsilon_k^2}  U_k .
\end{equation}
The Bogoliubov excitations energy is given by 
\begin{equation}\label {spec}
\varepsilon_k=\sqrt{E_k^2+ 2\mu_{0d}(\theta) E_k},
\end{equation}
where $\mu_{0d}(\theta)=n\lim\limits_{k\rightarrow 0} f({\bf k})$ is the zeroth order chemical potential.\\
Importantly, the spectrum (\ref {spec}) is independent of the random potential. 
This independence holds in fact only in zeroth order in perturbation theory; conversely, higher order calculations render the spectrum dependent on the random potential
due to the contribution of the anomalous terms (see below).
For $k\rightarrow 0$, the excitations are sound waves $\varepsilon_k=\hbar c_{sd} (\theta) k$, where $c_{sd} (\theta)=c_{s}\sqrt{1+\epsilon_{dd} (3\cos^2\theta-1)}$
with $c_{s}=\sqrt{gn/m}$ is the sound velocity without DDI.
Due to the anisotropy of the dipolar interaction, the sound velocity acquires a dependence on the propagation direction, which is fixed by the angle
$\theta$ between the propagation direction and the dipolar orientation. This angular dependence of the sound velocity has been confirmed experimentally \cite {bism}.

Therefore, the diagonal form of the Hamiltonian of the dirty dipolar Bose gas (\ref{ham1}) can be written as
\begin{equation}\label{DHami} 
\hat H = E+\sum\limits_{\vec k} \varepsilon_k\hat b^\dagger_{\bf k}\hat b_{\bf k},
\end{equation}
where $E=E_{0d}+\delta E+ E_R$, \\ $E_{0d}(\theta)=\mu_{0d}(\theta) N /2$ with $N$ being the total number of particles. 
\begin{equation}\label{Fenergy} 
\delta E=\frac{1}{2}\sum\limits_{\bf k} [\varepsilon_k -E_k-n f({\bf k})],
\end{equation}
is the ground-state energy correction due to qunatum fluctuations. 
\begin{equation}\label{Renergy} 
E_R=-\sum\limits_{\bf k} n\langle |U_k|^2\rangle \frac{ E_k}{\varepsilon_k^2} =-\sum\limits_{\bf k} n R_k \frac{ E_k}{\varepsilon_k^2},
\end{equation}
gives the correction to the ground-state energy due to the external random potential.

The noncondensed and the anomalous densities are  defined as  $\tilde{n}=\sum_{\bf k} \langle\hat a^\dagger_{\bf k}\hat a_{\bf k}\rangle$ 
and $\tilde{m}=\sum_{\bf k} \langle\hat a_{\bf k}\hat a_{-\bf k}\rangle$, respectively. Then invoking for the operators $a_{\bf k}$
the transformation (\ref {trans}), setting $\langle \hat b^\dagger_{\bf k}\hat b_{\bf k}\rangle=\delta_{\bf k' \bf k}N_k$ and putting the rest of the expectation values equal to zero,
where $N_k=[\exp(\varepsilon_k/T)-1]^{-1}$ are occupation numbers for the excitations.  
As we work in the thermodynamic limit, the sum over $k$ can be replaced by the integral $\sum_{\bf k}=V\int d^3k/(2\pi)^3$ 
and using the fact that $2N (x)+1= \coth (x/2)$, we obtain:
\begin{equation}\label {nor}
\tilde{n}=\frac{1}{2}\int \frac{d^3k} {(2\pi)^3} \frac{E_k+f({\bf k}) n} {\varepsilon_k}\left[\coth\left(\frac{\varepsilon_k}{2T}\right)-1\right]+n_R,
\end{equation}
and
\begin{equation}\label {anom}
\tilde{m}=-\frac{1}{2}\int \frac{d^3k} {(2\pi)^3} \frac{f({\bf k}) n} {\varepsilon_k}\coth\left(\frac{\varepsilon_k}{2T}\right)+n_R.
\end{equation}
The contribution of the random potential comes through the last terms in Eqs (\ref{nor}) and (\ref {anom}). These terms are defined as 
\begin{equation}\label {depdis}
n_R=\frac{1}{V}\sum\limits_{\bf k} \langle |\beta_{\bf k}|^2\rangle=n\int \frac{d^3k} {(2\pi)^3} R_k \frac{ E_k^2}{\varepsilon_k^4}.
\end{equation}
%First terms in Eqs.(\ref{nor}) and (\ref {anom}) are the zero-temperature contribution to the noncondensed $\tilde{n}_0$ and anomalous $\tilde{m}_0$ densities, respectively.
%Second terms represent the contribution of the so-called  thermal fluctuations and we denote them as  $\tilde{n}_T$ and $\tilde{m}_T$, respectively. 
%The total density of noncondensed particles is given by $\tilde{n}=\tilde{n}_0+\tilde{n}_T$ and the total anomalous density is written as $\tilde{m}=\tilde{m}_0+\tilde{m}_T$.\\
Expressions (\ref{nor}) and (\ref {anom}) must satisfy the equality 
\begin{align}\label {heis}
\tilde{n}_k(\tilde{n}_k+1)-|\tilde{m}_k|^2&= \frac{1}{4\, \text {sinh}^2\left(\varepsilon_k/2T\right)} \nonumber\\
&+n_R\left(\frac{E_k+2gn}{\varepsilon_k} \right) \text {coth} \left(\frac{\varepsilon_k}{2T}\right).
\end{align}
Equation (\ref {heis}) clearly shows that $\tilde{m}$ is larger than $\tilde{n}$ at low temperature irrespective of the presence of an external random potential or not. 
So the omission of the anomalous density in this situation is principally an unjustified approximation and wrong from the mathematical point of view \cite {boudj2011, boudj2012, boudj2015}.

\section{Fluctuations and thermodynamic quantities}
\label {zeroT}
To proceed further in practical calculations, we must specify the type of random potential. For this purpose, we
take the case of a spatially decaying disorder correlation $R({\bf r})$. 
Therefore, in what follows, we restrict ourselves to the case of a Gaussian correlation with the Fourier transform \cite {Mich, Axel1}
\begin{equation} \label{Disp}
R({\bf k}) = R e^{-\sigma^2k^2/2},
\end{equation}
where $R$ with dimension (energy) $^2$ $\times$ (length)$^3$ and $\sigma$ characterize the strength and the correlation length of the disorder, respectively.
Equation (\ref{Disp}) makes the macroscopic wave function of BEC not sensitive to disorder in and between pores, but instead
depends on the disorder averaged over the coherence length. Hence the ensemble-averaged system can become nearly uniform \cite {Mich}. 

\begin{figure}[htb1] 
\includegraphics[scale=0.9]{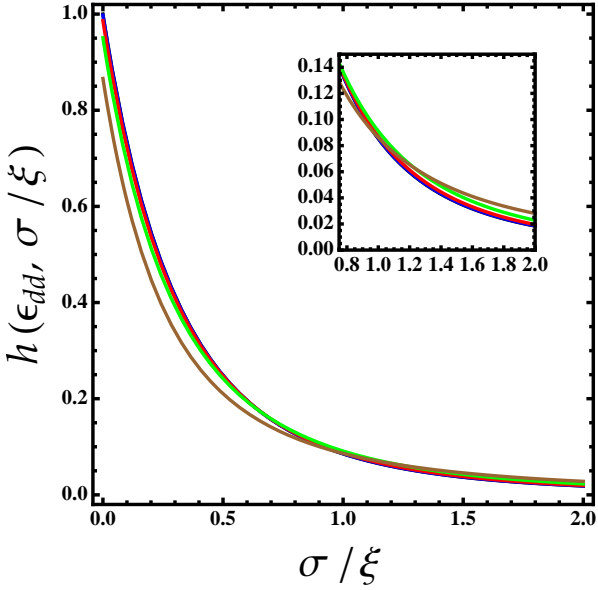}
\caption {(Color online) Behavior of the disorder function $h(\epsilon_{dd}, \sigma/\xi)$ from Eq.(\ref {func}),
 as a function of $\sigma/\xi$.  Black line: $\epsilon_{dd}=0$ (pure contact interaction), blue line: $\epsilon_{dd}=0.15$ (Cr atoms), red line: $\epsilon_{dd}=0.38$ (Er atoms), 
green line: $\epsilon_{dd}=0.7$ and brown line: $\epsilon_{dd}=1$.}
\label{dis}
\end{figure}

Substituting (\ref{Disp}) in Eq.(\ref{depdis}), we obtain for the condensate fluctuation due to the disordered potential 
\begin{equation} \label{depdis1}
{n_R}=\frac{m^2R}{8\pi^{3/2} \hbar^4} \sqrt{\frac{n}{a}} h(\epsilon_{dd}, \alpha),
\end{equation}
where 
\begin{equation} \label{func}
h(\epsilon_{dd}, \alpha)= \int_0^\pi \frac{d \theta \sin\theta {\cal F}(\alpha)}{2\sqrt{1+\epsilon_{dd} (3\cos^2\theta{_k}-1)}}, 
\end{equation}
is depicted in Fig.\ref{dis}.\\
The function ${\cal F}(\alpha)=e^{2\alpha} (4\alpha+1) \left[1-\text{erf}(\sqrt{2\alpha})\right]-2\sqrt{2\alpha/\pi}$ with
$\alpha=\sigma^2[1+\epsilon_{dd} (3\cos^2\theta-1)]/\xi^2$, has the following asymptotics for small $\alpha$: 
${\cal F}(\alpha)=1-4 \sqrt{2\alpha/\pi} +6\alpha -(32/3) \alpha \sqrt{2\alpha/\pi} +10 \alpha ^2+O\left(\alpha ^{5/2}\right)$.
Equation.(\ref{depdis1}) is in good agreement with that obtained using the mean field theory \cite {Axel1}. 

We observe from Fig.\ref{dis} that for a small disorder correlation length $\sigma<\xi $, the contribution of the DDI on the disorder fluctuation is not important whereas
in the case of $\sigma>\xi $, the DDIs tend to enhance the disorder fluctuation.\\
On the other hand, for $\sigma/\xi\rightarrow 0$, we get from Eq.(\ref{func}) that $h(\epsilon_{dd}, 0) ={\cal Q}_{-1}(\epsilon_{dd})$. 
Thus, the disorder fluctuation (\ref{depdis1}) becomes identical to that obtained in 3D dipolar BEC with delta-correlated disorder \cite {Axel3} 
$n_R=(m^2R/8\pi^{3/2} \hbar^4) \sqrt{n/a} {\cal Q}_{-1}(\epsilon_{dd})$,
where the contribution of the DDI is expressed by the functions ${\cal Q}_j(\epsilon_{dd})=(1-\epsilon_{dd})^{j/2} {}_2 \!F_1\left(-\frac{j}{2},\frac{1}{2};\frac{3}{2};\frac{3\epsilon_{dd}}{\epsilon_{dd}-1}\right)$, where ${}_2 \!F_1$ is the hypergeometric function. Note that functions ${\cal Q}_j(\epsilon_{dd})$ attain their maximal values for $\epsilon_{dd} \approx 1$ 
and become imaginary for $\epsilon_{dd}>1$ \cite {lime, boudj2015}.\\
For $\sigma/\xi\rightarrow 0$ and $\epsilon_{dd}=0$, we read off from Eq.(\ref{func}) that one obtains $h(\epsilon_{dd}, \alpha) \rightarrow 1$. 
Therefore, we should reproduce the Huang and Meng result \cite{Huang} for the disorder fluctuation in this limit. 

%Another important consequence is that  when $a$ vanishes, $n_R$ becomes infinite. This means that the system would collapse if there were no repulsive interactions between particles. 

Upon calculating the integral in Eq.(\ref{nor}), we  get for the noncondensate depletion 
\begin{align}\label {dep}
\frac{\tilde{n}}{n}=&\frac{ 8}{3} \sqrt{\frac{ n a^3}{\pi}} {\cal Q}_3(\epsilon_{dd})+\frac{2}{3}\sqrt{\frac{n a^3}{\pi}} \left(\frac{\pi T}{gn}\right)^{2}{\cal Q}_{-1}(\epsilon_{dd}) \nonumber\\
&+2\pi R' \sqrt{\frac{ n a^3}{\pi}} h(\epsilon_{dd}, \alpha),
\end{align}
where $R'=R/g^2n$ is a dimensionless disorder strength.
The condensed fraction can be calculated employing  $n_c/n=1-\tilde{n}/n$.

The integral in Eq.(\ref{anom}) is ultraviolet divergent. This divergence is well-known to be unphysical, since it is caused by
the usage of the contact interaction potential as we have mentioned in Sec.\ref{model}. A general way of treating such integrals is as follows. First, one restricts to
asymptotically weak coupling and introduces the Beliaev-type second order coupling constant \cite {boudj2015}
\begin{equation}\label {cc}
f_R ({\bf k})=f({\bf k}) -\frac{m}{\hbar^2}\int \frac {d^3q}{(2\pi)^3} \frac{f(-{\bf q}) f({\bf q})}{2E_q}.
\end{equation}
After the subtraction of the ultraviolet divergent part, the anomalous fraction turns out to be given as
\begin{align}\label {anom1}
\frac{\tilde{m}}{n}=&8\sqrt{\frac{ n a^3}{\pi}} {\cal Q}_3(\epsilon_{dd})-\frac{2}{3}\sqrt{\frac{n a^3}{\pi}} \left(\frac{\pi T}{gn}\right)^{2}{\cal Q}_{-1}(\epsilon_{dd})\nonumber\\
&+2\pi R' \sqrt{\frac{ n a^3}{\pi}} h(\epsilon_{dd}, \alpha).
\end{align}
The leading terms in Eqs.(\ref{dep}) and (\ref{anom1}) represent the qunatum fluctuation\cite {boudj2015}. 
The subleading terms which represent the thermal fluctuation\cite {boudj2015}, are calculated at temperatures $T\ll gn $, 
where the main contribution to integrals (\ref{nor}) and (\ref{anom}) comes from the region of small momenta ($\varepsilon_k=\hbar c_{sd} k$).
The situation is quite different at higher temperatures i.e. $T\gg gn$, where the main contribution to integrals (\ref{nor}) and (\ref{anom}) comes from the single particle excitations. 
Hence, the thermal contribution of $\tilde{n}$ becomes identical to the density of noncondensed atoms in an ideal Bose gas \cite{boudj2015}, while
the thermal contribution of $\tilde{m}$ tends to zero since the gas is completely thermalized in this range of temperature \cite{Griffin, boudj2011, boudj2015}.
The last terms in (\ref{dep}) and (\ref{anom1}) describe the effect of disorder on the noncondensed and on the anomalous densities.

Equation (\ref {anom1}) clearly shows that at zero temperature, the anomalous density is three times larger than the noncondensed density 
for any range of the dipolar interaction as well as for any value of the strength and the correlation length of the disorder as it has been anticipated above. 
Moreover, $\tilde{m}$ changes its sign with increasing temperature in agreement with uniform Bose gas with a pure contact interaction \cite {boudj2015}.
For $\epsilon_{dd}=0$, ${\cal Q}_j(\epsilon_{dd})=1$ and thus, Eqs.(\ref {dep}) and (\ref{anom1}) reproduce the short-range interaction results.
Furthermore, the DDI enhances the condensate depletion and the anomalous fraction for increasing $\epsilon_{dd}$.

The condensate fluctuations manifest into the second-order correlation function as \cite {Glaub}
\begin{align}\label {correl}
G^{(2)}(r)&=\langle\hat \psi^\dagger(r)\hat \psi^\dagger(r)\hat \psi(r) \hat\psi (r)\rangle \nonumber \\
 &=n_{c}^2+\tilde{m}^2+2\tilde{n}^2+4\tilde{n}n_c+2\tilde{m}n_c.
\end{align}
Equation (\ref{correl}) is obtained using Wick’s theorem. Inserting then Eqs.(\ref{depdis1}), (\ref {dep}) and (\ref {anom1}) into (\ref{correl}), we obtain
\begin{equation}\label {G}
\frac{G^{(2)}}{n^2}=1+\frac{64}{3}\sqrt{\frac{n a^3 }{\pi}} {\cal Q}_3(\epsilon_{dd})+8\pi R' \sqrt{\frac{ n a^3}{\pi}} h(\epsilon_{dd}, \alpha).
\end{equation}
This equation is accurate to the first order in  $\tilde{n}/n_c$ and $\tilde{m}/n_c$ and shows how the correlation function depends on the interaction parameter $\epsilon_{dd}$ and
 the disorder length $\sigma$.

\begin{figure}[htb1] 
\includegraphics[scale=0.9]{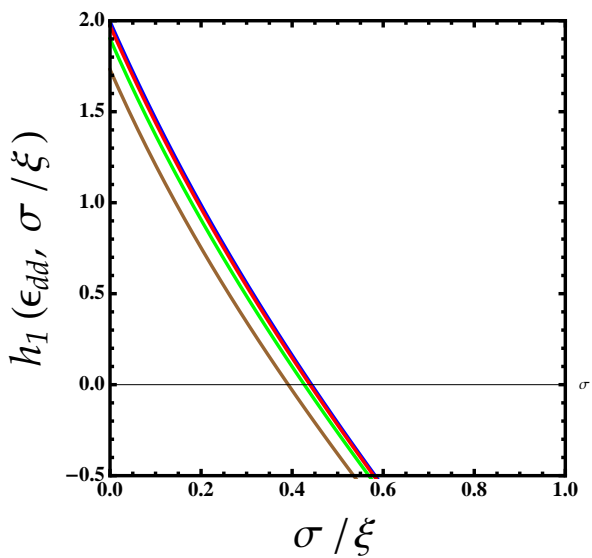}
\caption {Behavior of the disorder energy function $h_1(\epsilon_{dd}, \sigma/\xi)$, from Eq. (\ref{func1})  as a function of $\sigma/\xi$ 
 for same values of $\epsilon_{dd}$ as in Fig.\ref{dis}.}
\label{dis1}
\end{figure}

The energy shift due to the interaction and the quantum fluctuations (\ref {Fenergy}) are ultraviolet divergent. The difficulty is
overcome if one takes into account the second-order correction to the coupling constant (\ref {cc}). A straightforward calculation yields\cite {lime, boudj2015}
\begin{equation}\label {energ} 
\delta E=\frac{64}{15}Vgn^2\sqrt{\frac{n a^3}{\pi}} {\cal Q}_5(\epsilon_{dd}).
\end{equation} 
However, the energy shift (\ref {Renergy}) due to the external random potential (\ref {Disp}) is not divergent  and it can be evaluated as 
\begin{equation}\label {Reng}
\frac{E_R}{E_0}=16 \pi R' \sqrt{\frac{n a^3}{\pi}} h_1(\epsilon_{dd}, \alpha),
\end{equation}
where  $E_0=N gn/2$, and 
\begin{equation} \label{func1}
h_1(\epsilon_{dd}, \alpha)= \frac{1}{2}\int_0^\pi d\theta \sin \sqrt{1+\epsilon_{dd} (3\cos^2\theta{_k}-1)} {\cal F}_1(\alpha),
\end{equation}
is displayed in Fig.\ref{dis1}.\\
The function ${\cal F}_1(\alpha)=e^{2\alpha} [1-\text{erf}(\sqrt{2\alpha})] -\sqrt{1/2\pi\alpha}$ has the asymptotics behavior for small $\alpha$:
${\cal F}_1(\alpha)=1- (4+\pi ) \sqrt{\alpha /2 \pi}+2 \alpha -(8/3)\sqrt{ 2/\pi } \alpha ^{3/2}+2 \alpha ^2+O\left(\alpha ^{5/2}\right)$.\\
As is seen from Fig.\ref{dis1} that for $\sigma <2\xi$, the energy correction due to the disorder effect (\ref {Reng}) is negative which leads to lower the total energy of the system.
Note that this result still valid for any value of $\epsilon_{dd} <1$. Another important remark is that the energy decreases with increasing $\epsilon_{dd}$.\\
For a condensate with a pure contact interaction (${\cal Q}_5(\epsilon_{dd}=0)=1$) and in the absence of disordered potential ($R=0$),
the obtained energy excellently agrees with the seminal Lee-Huang-Yang quantum corrected ground state energy \cite{LHY}.\\
For, $\sigma /\xi \rightarrow 0$, the energy shift due to the external random potential (\ref {Renergy}) becomes ultraviolet divergent. 
Again, by introducing the renormalized coupling constant (\ref{cc}) one gets:
$E_R/E_0=16 \pi R' \sqrt{n a^3/\pi} {\cal Q}_1(\epsilon_{dd})$ which well coincides with the result obtained with delta-correlated disorder of Ref \cite {Axel3}.

The chemical potential can be obtained easily through $\mu= \partial E/\partial N$ or via $\mu=\mu_0+\delta \mu$, where
$\delta \mu=\sum_{\bf k} f({\bf k} ) [v_k(v_k-u_k)]=\sum_{\bf k} f({\bf k} ) (\tilde{n}+\tilde{m})$ \cite{boudj2015, boudj2012, Abdougora}. 
One can also calculate the shift of the sound velocity utilizing $mc_s^2 = n \partial \mu/\partial n $ \cite {Lev}.

The Bogoliubov approach assumes that fluctuations should be small. We thus conclude from Eqs. (\ref {dep}) and (\ref {anom1}) that at $T = 0$, the validity of the 
Bogoliubov theory requires inequalities $\sqrt{n a^3}  {\cal Q}_{3} (\epsilon_{dd})\ll 1$ and $ R' \sqrt{n a^3} h(\epsilon_{dd}, \alpha)\ll 1$.
For $R'=0$, this parameter differs only by the factor ${\cal Q}_3(\epsilon_{dd})$ from the universal small parameter of the theory, $\sqrt{na^3}\ll 1$, in the absence of DDI. 
At $T \ll gn$, the Bogoliubov theory requires the condition $(T/gn) \sqrt{n a^3}  {\cal Q}_{-1}(\epsilon_{dd})\ll 1$. The appearance of the extra factor ($T/gn $) 
originates from the thermal fluctuations corrections.

\section{superfluid fraction} \label {superfluid}
The superfluid fraction $n_s/n$ can be found from the normal fraction $n_n/n$ which is determined by the transverse current-current
correlator $n_s/n =1-n_n/n$. We apply a Galilean boost with the total momentum of the moving system ${\bf P}=mv (n {\bf v_s}+n_n {\bf v_n})$, where 
${\bf v_s}$ denotes the superfluid velocity and and ${\bf v_n}={\bf u}-{\bf v_s}$ is the normal fluid  velocity with ${\bf u}$ being a boost velocity \cite{Axel2}. 
Using (\ref{heis}), the superfluid fraction is then written 
\begin{equation} \label{sup1}
 \begin{split}
 \frac{n_s^{ij}}{n}= \delta_{ij}-4\int \frac{d^3k}{(2\pi)^3} \frac{\hbar^2}{2m} \frac{n R_k k_ik_j}{E_k[E_k-2n f({\bf k})]^2} \\ 
-\frac{2}{Tn}\int \frac{d^3k}{(2\pi)^3} \left[\frac{\hbar^2}{2m} \frac{k_i k_j}{4\, \text {sinh}^2 (\varepsilon_k/2T)}\right]. 
\end{split}
\end{equation}
It is worth noticing that if in expression (\ref{sup1}) $\tilde {m}$ were omitted, then the superfluid fraction would be divergent signaling 
the importance of the anomalous density  on the occurrence of the superfluidity in Bose gases \cite{boudj2015, Yuk}.

\begin{figure}[htb1] 
\centerline{
\includegraphics[scale=0.9]{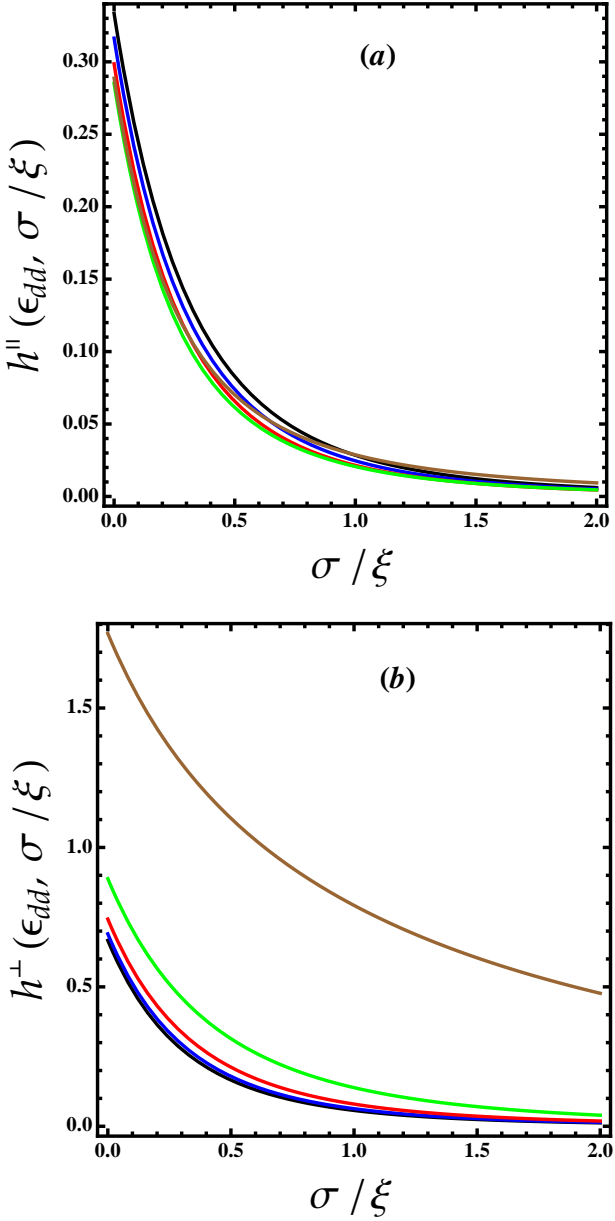}}
\caption {(Color online) Behavior of the disorder functions $h^{\parallel}(\epsilon_{dd}, \sigma/\xi)$ (a) and $h^{\perp}(\epsilon_{dd}, \sigma/\xi)$ (b),
 as a function of $\sigma/\xi$ for same values of $\epsilon_{dd}$ as in Fig.\ref{dis}.}
\label{disP}
\end{figure}

\begin{figure}[htb1] 
\centerline{
\includegraphics[scale=0.9]{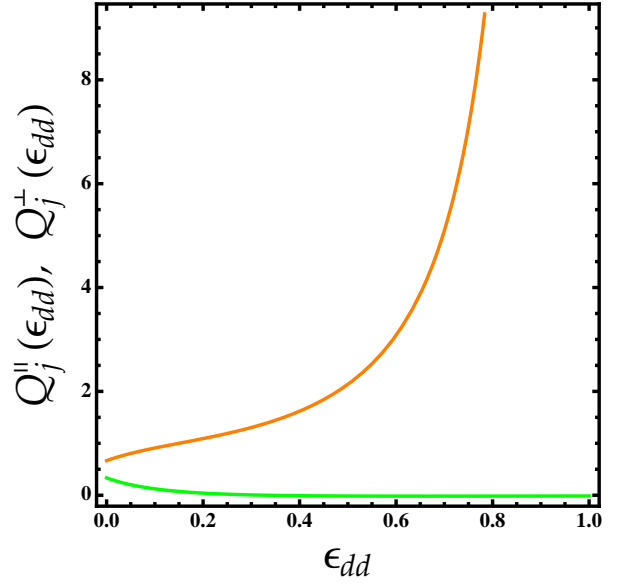}}
\caption {(Color online) Behavior of thermal functions ${\cal Q}_j^{\parallel}$ (Green line) and ${\cal Q}_j^{\perp}$ (Orange line), as a function of $\epsilon_{dd}$.}
\label{therm}
\end{figure}

Equation. (\ref{sup1})  yields a superfluid density that depends on the direction of the superfluid motion with respect to
the orientation of the dipoles. In the parallel direction and at low temperatures where $\varepsilon_k=\hbar c_{sd} k$, the superfluid fraction reads

\begin{equation}\label {supflui1}
\frac{n_s^{\parallel}}{n} =1-4\pi R' \sqrt{\frac{n a^3}{\pi}} h^{\parallel}(\epsilon_{dd}, \alpha)-\frac{2\pi^2T^4}{45 mn\hbar^3 c_s^5} {\cal Q}_{-5}^{\parallel}(\epsilon_{dd}),
\end{equation}
where the function
\begin{equation} \label{funcP}
h^{\parallel}(\epsilon_{dd}, \alpha)= \int_0^\pi d\theta \frac{ \sin\theta\cos^2\theta {\cal F} (\alpha)}{2\sqrt{1+\epsilon_{dd} (3\cos^2\theta-1)}}, 
\end{equation}
is decreasing with increasing $\epsilon_{dd}$ and vanishing for large $\sigma/\xi$ as is depicted in Fig.\ref{disP}.a. And the functions
${\cal Q}_j^{\parallel} (\epsilon_{dd})=\frac{1}{3}(1-\epsilon_{dd})^{j/2} {}_2\!F_1\left(-\frac{j}{2},\frac{5}{2};\frac{3}{2};\frac{3\epsilon_{dd}}{\epsilon_{dd}-1}\right)$, 
have the properties ${\cal Q}_j^{\parallel} (\epsilon_{dd}=0)=1/3$ and become imaginary for $\epsilon_{dd}>1$ (see Fig.\ref{therm}).
Therefore, Eq.(\ref{supflui1}) reveals that DDI effects are more significant for condensate fracion (\ref {dep}) than for the parallel superfluid fraction. 

Again at low temperatures,  the perpendicular direction of the superfluid fraction (\ref{sup1}) takes the form
\begin{equation}\label {supflui2}
\frac{n_s^{\perp}}{n} =1-2\pi R' \sqrt{\frac{n a^3}{\pi}} h^{\perp}(\epsilon_{dd}, \alpha)-\frac{\pi^2T^4}{45 mn\hbar^3 c_s^5} {\cal Q}_{-5}^{\perp}(\epsilon_{dd}),
\end{equation}
where the functions 
\begin{align} \label{funcP2}
h^{\perp}(\epsilon_{dd}, \alpha)=& \int_0^\pi d\theta \frac{ \sin\theta(1-\cos^2\theta) {\cal F}(\alpha)}{2\sqrt{1+\epsilon_{dd} (3\cos^2\theta-1)}} \nonumber\\
&=h(\epsilon_{dd}, \alpha)-h^{\parallel}(\epsilon_{dd}, \alpha), 
\end{align}
and ${\cal Q}_j^{\perp}(\epsilon_{dd})= {\cal Q}_j(\epsilon_{dd})-{\cal Q}_j^{\parallel} (\epsilon_{dd})$, are displayed in Figs.(\ref{disP}.b), and (\ref{therm}), respectively.

Expressions (\ref{supflui1}) and (\ref{supflui2}) constitute a natural extension of those obtained in \cite{Axel1} since they contain the temperature correction (third terms). 
At $T\gg ng$, it is evident that both thermal terms of $n_s$ coincide with the noncondensed density of an ideal Bose gas.
Figure (\ref{therm}) shows that the thermal contribution of  $n_s^{\perp}$ is smaller than that of $n_s^{\parallel}$  for $\epsilon_{dd}\leq 0.5$, 
while the situation is inverted for  $\epsilon_{dd}> 0.5$.\\
For $\sigma/\xi\rightarrow 0$ and $\epsilon_{dd}=0$, both components of the superfluid fraction (\ref{supflui1}) and (\ref{supflui2}) reduce to $n_s/n=1-4n_R/3n$,
which well recove earlier results of Refs \cite {Huang, Gior, Lopa} for isotropic contact interaction. 
For $\sigma/\xi\rightarrow 0$, we have $h^{\parallel}(\epsilon_{dd}, 0) ={\cal Q}^{\parallel}_{-1}(\epsilon_{dd})$ and $h^{\perp}(\epsilon_{dd}, 0) ={\cal Q}^{\perp}_{-1}(\epsilon_{dd})$. 
Consequently, the disorder correction to superfluid fraction (\ref{depdis1}) becomes identical to that obtained in 3D dipolar BEC with delta-correlated disorder \cite {Axel3}. 
We should stress also that for increasing $\epsilon_{dd}$,  $h^{\parallel}(\epsilon_{dd}, \alpha)$ decreases, whereas $h(\epsilon_{dd}, \alpha)$ 
increases for fixed $\sigma/\xi$.
Therefore, this reveals that there exists a critical value $\epsilon^c_{dd}$ beyond which the system has the surprising property that
the disorder-induced depletion of the parallel superfluid density is smaller than the condensate depletion even at $T=0$.
This can be attributed to the fact that the localized particles cannot contribute to superfluidity and, hence, form
obstacles for the superfluid flow. 
For a large disorder correlation length i.e. $\sigma \gg \xi$, $\epsilon^c_{dd}$ decreases indicating that the locally condensed particles are
localized in the respective minima for the disorder potential for a finite localization time \cite {Axel4}.

The superfluid fraction can be either larger or smaller than the condensate fraction $n_c/n$, depending on temperature, the
strength of interactions, and on the strength of disorder. Increasing $R'$ leads to the simultaneous
disappearance of the superfluid and condensate fractions. So for any value of $n a^3$ and of $\epsilon_{dd}$ there exists a critical strength of disorder 
\begin{equation}\label {RR}
R^{\parallel}_c= \frac{4}{\pi} \frac{{\cal Q}_{3}(\epsilon_{dd})}{h^{\parallel}(\epsilon_{dd}, \alpha)}, \,\,\,\,\, 
R^{\perp}_c= \frac{2}{\pi} \frac{{\cal Q}_{3}(\epsilon_{dd})}{h^{\perp}(\epsilon_{dd}, \alpha)},
\end{equation}
for which $n_s^{\parallel}/n < n_c/n$ and $n_s^{\perp}/n < n_c/n$.\\
When $\sigma/\xi\rightarrow 0$, $R^{\parallel}_c= (4/\pi) [{\cal Q}_{3}(\epsilon_{dd})/ {\cal Q}^{\parallel}_{-1}(\epsilon_{dd})]$ 
and $R^{\perp}_c= (2/\pi) [{\cal Q}_{3}(\epsilon_{dd})/ {\cal Q}^{\perp}_{-1}(\epsilon_{dd})]$ .
In the case of Er atoms ($\epsilon_{dd}=0.38$), $R^{\parallel}_c \approx 6.74$ and $R^{\perp}_c \approx 0.78$. 
For Cr atoms ($\epsilon_{dd}=0.15$), $R^{\parallel}_c \approx 4.96$ and $R^{\perp}_c \approx 0.86$. This clearly shows that $R^{\parallel}_c$ is decreasing with $\epsilon_{dd}$,
while  $R^{\perp}_c$ is increasing with $\epsilon_{dd}$.\\
Therefore, the  Bogoliubov approach should satisfy the condition:$R'<R_c$.
However, it is not clear whether these results are still valid for $R'>R_c$ in a range of densities where the difference between
$n_s/n$ and $n_c/n$ can be significant and hence, the system yields a transition to a new quantum regime. 
The response to these questions requires either a non-perturbative scheme or numerical Quantum Monte Carlo simulation, 
which are out of the scope of the present work.

\section{Trapped dipolar BEC with weak disorder} \label{Trap}

In this section, we discuss the case of a harmonically trapped dipolar Bose gas with weak disorder.
Let us start by writing the hydrodynamic equations.  
These equations can be derived  by factorizing the condensate wave function according to the Madelung transformation
$\phi({\bf r},t)=\sqrt{n({\bf r},t)} \exp (-i S({\bf r},t))$,
where $S$ is the phase of the order parameter, it is  a real quantity related to the superfluid  velocity by
$v=\hbar/m {\bf \nabla} S$. 
The continuity and Euler-like equations read, respectively:
\begin{align} \label{hydo}
&\frac{\partial n}{\partial t} +{\bf \nabla}.(n v)=0,\\
&m\frac{\partial v}{\partial t} =-{\bf \nabla} \left [-\frac{\hbar^2}{2m} \frac{\Delta \sqrt{n}} {\sqrt{n}} +\frac{1}{2} m v^2+{\cal U}+gn+\Phi_{dd}\right],
\end{align}
where $\Delta \sqrt{n}/ \sqrt{n}$ stands for the quantum pressure, $\Phi_{dd}=\int d\mathbf{r}^\prime\, V_d (\mathbf{r}-\mathbf{r^\prime})|\phi (\mathbf{r^\prime})|^2 $
and ${\cal U}({\bf r})=U_{trap}({\bf r})+U({\bf r})$ with $U_{trap}({\bf r})=m (\omega_x^2 x^2+\omega_y^2 y^2+\omega_z^2 z^2)/2$ being the trapping potential 
and $\omega_j$ are the trapping frequencies.\\
In a nonstationary situation, we consider small oscillations (low density) for the density around its static solution in the form
$n=n_0+\delta n$, where  $\delta n/n_0\ll1$. \\
Shifting the phase by $-\mu t/\hbar$, we then linearize Eq.(\ref {hydo}) with respect to $\delta n$ and ${\bf \nabla} S$ around the stationary solution.  
The zero-order terms give for the chemical potential:
\begin{equation}  \label{chim}
\mu=-\frac{\hbar^2}{2m} \frac{\Delta \sqrt{n_0}} {\sqrt{n_0}}+{\cal U}({\bf r})+gn_0({\bf r})+\Phi_{dd}.
\end{equation}
Expanding the density and phase in the basis of the excitations
$\hat S({\bf r})=\left(-i/2\sqrt{n_0({\bf r})}\right) \sum_k [f_k^+({\bf r}) \hat {b}_k-h.c.]$ and $\delta \hat n ({\bf r})=\sqrt{n_0({\bf r})} \sum_k [f_k^-({\bf r}) \hat {b}_k+h.c.]$,
one then finds that $f_k^{\pm}=u_k\pm v_k$ obey the Bogoliubov-de Gennes equations (BdGE):
\begin{align}   \label{BdG}
&\left[-\frac{\hbar^2}{2m} \Delta+ {\cal U} ({\bf r}) +gn_0({\bf r})+\Phi_{dd}-\mu \right] f_k^+({\bf r})=\varepsilon_k f_k^-({\bf r}),\\
&\left[-\frac{\hbar^2}{2m} \Delta+ {\cal U}({\bf r}) +3gn_0({\bf r})+\Phi_{dd}-\mu \right] f_k^-({\bf r})=\varepsilon_k f_k^+({\bf r}). \nonumber
\end{align}
Equations (\ref{chim}) and (\ref{BdG}) form a complete set to calculate the 
ground state (BEC) and excitations of the dirty dipolar Bose gas, from which one can compute all the properties of finite
temperature, collective modes or time-dependent dipolar BEC with random potential.

To derive analytical expressions for the physical quantities of interest such as the condensate depletion, the equation of state, the ground state energy and so on,
one should use the local density approximation (LDA)  (for more details on the applicability of this approximation on long-range interactions, see e.g. \cite {lime}).
The LDA or semiclassical approximation is applicable when the external potential is sufficiently smooth, and requires that $f_k^{\pm}({\bf r})$ 
are slowly varying functions of the position as well as $U({\bf r})$ is assumed to be a weak random potential.
Employing  the LDA, the BdGE (\ref{BdG}) take a simple algebraic form as in the homogeneous case. Therefore,
the local Bogoliubov spectrum can be obtained in the usual way and reads $\varepsilon_k({\bf r})=\sqrt{E_k^2+ 2 n_0({\bf r}) f_k E_k}$. 
Then, we obtain for the condensate depletion, the anomalous fraction and the ground state energy the following relations, respectively:
\begin{align}\label {depLADA}
\frac{\tilde{n}({\bf r})}{n_0({\bf r})}=\frac{ 8}{3} \sqrt{\frac{ n_0({\bf r}) a^3}{\pi}} {\cal Q}_3(\epsilon_{dd})+2\pi R' \sqrt{\frac{ n_0({\bf r}) a^3}{\pi}} h(\epsilon_{dd}, \alpha)+\cdots,
\end{align}
\begin{align}\label {anom1LDA}
\frac{\tilde{m}({\bf r})}{n_0({\bf r})}=8\sqrt{\frac{ n_0({\bf r}) a^3}{\pi}} {\cal Q}_3(\epsilon_{dd})+2\pi R' \sqrt{\frac{ n_0({\bf r}) a^3}{\pi}} h(\epsilon_{dd}, \alpha)+\cdots,
\end{align}
and
\begin{equation}\label {Reng}
\frac{E_R}{E_0}=16 \pi R' \sqrt{\frac{n_0({\bf r}) a^3}{\pi}} h_1(\epsilon_{dd}, \alpha),
\end{equation}
where $E_0=N gn_0({\bf r})/2$.\\
In the regime where the number of particles is very large (Thomas-Fermi limit), the density $n_0$ remains extended (delocalized) for a weak $U$:
$n_0({\bf r})=(\mu-{\cal U}({\bf r})-\Phi_{dd})/g$. 
Equivalently, one can also use Eqs.(\ref{depLADA}) and (\ref{anom1LDA}) to calculate the corrections to the equation of state and the local superfluid density. 
Importantly, we see that the disorder corrections remain anisotropic as in the homogeneous case.
% which is indeed an advantage of the LDA.
%The fact that these corrections remain the same as in the homogeneous case, this an artifact of the LDA. 

\section{Conclusions}\label{conc}

In this paper, we have studied the properties of a dipolar Bose gas in the presence of a weak random potential with a Gaussian correlation at finite temperature. 
Within the Bogoliubov approach, we have calculated the condensate fluctuation due to disorder, 
as well as the corresponding corrections to the noncondensed and the anomalous densities, second order correlation function, the ground state energy and the superfluid fraction
in the homogeneous case. 
We have pointed out that the interplay of the DDI and the external random potential makes both the BEC and the superfluidity anisotropic.
In particular, we have discussed the consequences of the superfluid density, which becomes a tensorial quantity
as a linear response to the moving disorder. 
We have found that the presence of the DDI causes a weak anisotropy in the parallel component of the superfluid density while the perpendicular component becomes strongly  anisotropic.
We have also demonstrated that for a strong disorder strength the system introduces an unusual quantum regime where the superfluid fraction is smaller than the condensate fraction.
In addition, we have reproduced the expression of the condensate fluctuations and thermodynamic quantities
obtained earlier within the treatment of Huang and Meng \cite {Huang} in the absence of the DDI and those obtained recently in Ref \cite {Axel3} in the limit $\sigma/\xi\rightarrow 0$. 
Furthermore, we have discussed the validity criterion of the Bogoliubov approach in a disordered dipolar BEC at finite temperatures. 
By means of the LDA, we have generalized our results to the case of a harmonically trapped gas.
These results could be directly applied to check how quantum and disorder
fluctuations can alter the time-of-flight expansion of a dirty-trapped dipolar Bose gas.

Promising candidates for the experimental realization of such dirty dipolar BECs are atomic species with highly magnetic dipolar interaction such as 
Dy (magnetic moment 10$\mu_B$) \cite{Lu1} or polar heteronuclear molecules such as KRb (magnetic moment 0.6 Debye)\cite {Aik2}. 

Finally, an important step for future theoretical studies is to analyze the effects of an external random potential in a dilute 2D dipolar Bose gas near the roton minimum. 
In 1D case, the investigation of a dirty dipolar BEC in an optical lattice is expected to exhibit even richer
insights about the stability of the superglass state in Bose gases \cite{Zamp}.

%Finally, let us discuss some possible extensions of our work, 

\section{Acknowledgements}
We are grateful to Axel Pelster for valuable discussions.

\end{document}